\documentclass[conference]{IEEEtran}

\usepackage{graphicx}
\usepackage{syntax}
\usepackage{fancyhdr}

\begin{document}
\title{Dynamically Allocated Memory Verification in Object-Oriented Programs using Prolog}

\author{\IEEEauthorblockN{Ren\'{e} Haberland \qquad Sergey Ivanovskiy}
\IEEEauthorblockA{Department of Software Engineering and Computer Applications\\
Saint Petersburg Electrotechnical University "LETI"\\
Saint Petersburg, Russia }}

\maketitle

\thispagestyle{fancy}
\cfoot{This work is licensed under the Creative Commons Attribution License.}

\begin{abstract}
A Prolog-based framework for fully automated verification currently under development for heap-based object-oriented
data is introduced. Dynamically allocated issues are discussed, recent approaches and criteria
are analysed. The architecture and its components are introduced by example. Finally, propositions to further and related work are given.
\end{abstract}

%
\IEEEpeerreviewmaketitle

\section{Introduction}
The main interest of this work is dedicated to the correctness of a program according to its memory consumption behaviour. It may, however, also be extended to performance considerations based on results coming from the dynamic memory verification, particularly but not only during the alias analysis \cite{lit16,lit3} or during the garbage collection phase, for instance. A Prolog-based verifier is suggested.

 The structure of this paper is to give a short overview on current approaches in section 2. Section 3 presents the languages being used as programming and specification languages, and introduces the overall architecture. The architecture is designed to be open. Currently the project is under progress, the final section gives an outlook on related and future work.\\
 
\cite{lit9} discusses why the aliasing issue has still not been solved yet. \cite{lit15} provides a more detailed mark on the issues related to aliasing issues in a commercial Unix-environment. Despite its age \cite{lit15} and its partial closeness because of the commercial background, it is still often cited in very recent publications and numerous technical reports on the same topic, and in open-source projects, and surprisingly enough most of the issues found initially are still there almost with no changes. One key aspect that makes aliasing such a hard issue is that its local changes in a program listing may effect other regions unexpectedly --– but at the same time this is its strength, since no additional copying is required. \cite{lit15} had particularly analysed previously fixed bugs for a long-term period over last releases and found out - according to the bug distribution over time - that undesired memory behaviour is one of the most expensive bug reasons in terms of time and efforts to locate and 
fix.\\

Object-orientation \cite{lit1}, based on concepts such as encapsulation, polymorphism and inheritance, has been one of the most successful and widely adapted programming paradigms by now for a long time in industry, hence its combination with pointer structures remains a relevant research task up to date \cite{lit13}.\\

Prolog \cite{lit27} is considered for program verification for several reasons. First, it is a logical and declarative programming language which offers a high abstraction in writing Horn-clauses as they correspond with defined axioms and rules. A proof tree occasionally insists on a back-tracking strategy, which Prolog supports for free as one of its core-language features. The hope is Prolog's \textit{generate-and-test} goal strategy \cite{lit27} may be found useful in simplifying and abstracting a proof significantly. Second, programs and internal states can be represented as terms. Terms can be easily processed in Prolog. The hope is, abduction and general symbolic term evaluation will allow generating lemmas more efficiently and make the reasoning terminate and terminate earlier. As a previous successful verification attempt, \cite{lit21} shall be noted. The authors solved a fairly hard problem from mathematical numerics using Prolog elegantly and straight. Since proofs are rule-centric, a 
proof contradiction will eventually help generating counter-examples easily by simply matching terms from the memory and from a rule or axiom.\\

\textbf{Example1} --– memory leak
\begin{verbatim}
MyClass object1=new MyClass();
...
object1=new MyClass();
\end{verbatim}

\textbf{Example2} --- unachievable memory
\begin{verbatim}
// object1 has been created
MyClass object2=new MyClass();
object2.ref=object1;
\end{verbatim}

\textbf{Example3} --– invalid memory access
\begin{verbatim}
// object1.ref==null
value = (object1.ref).attribute1;
\end{verbatim}

\textbf{Example4} --– data structure with cycle
\begin{verbatim}
object1.next=object1;
...
root=object1;
while(root.next!=null){
  printf(“%d”, object.data);
  root=root.next;
}
\end{verbatim}

Example 1 demonstrates the case where a fresh memory region is allocated, and without freeing it, allocates it again, which may cause the previously created region becoming unachievable.  The second example demonstrates where \texttt{object2} is linked to an occupied \texttt{object1}, but \texttt{object2} itself remains unused. A very common problem in practice might be the third example, when an object reference is not set, but later referenced causing either an abnormal runtime failure or continues execution, which might be even worse in realistic scenarios because the further program execution becomes totally unpredictable with invalid value settings. The forth example might not immediately be seen as a problem, but if, for whatever reason, there is a cycle in \texttt{root} the program will not terminate. Apart from direct consequences like crashes or non-termination, one more side effect is there are spontaneous allocations/deallocations taking place on runtime which may eventually become a performance 
bottleneck.

\section{Current Approaches}

One approach to verify correctness of dynamic memory is to get the program run and to record all memory cells that will be referenced and allocated/deallocated. This is what the \textit{valgrind} tool \cite{lit29} does. This open-source tool requires on compilation guarding memory-checking code is injected to the assembly code. Not only that the enhanced program runs with huge delays, the general problem underneath this approach as well as SAFECode \cite{lit25}, which on runtime checks whether programmer-inserted assertions are fulfilled, is that only a small subset of all possible execution paths can be tested and that it requires additional code is inserted. For this reason, only static approaches are considered further that analyse the incoming program listing prior to running it.\\

In order to address the problems mentioned in the introduction part, several approaches exist: (i) Shape-based Analysis \cite{lit26,lit20}, (ii) Separating heap. For sake of completeness a heap-free alternative proposed by (iii) Tufte and Talpin \cite{lit28} and Meyer \cite{lit14} shall be mentioned, who both appeal to a stack-based approach, if any possible, to avoid expensive heap allocation and deallocation operations. Automatic handling of stack-based locations is essential for both, where the stack sizes are determined during compilation. Thus, control passing which happens during a call will allow to allocate objects almost for \textit{free}, since a stack frame needs to be created in any case, and no more expensive heap operations are needed in fact, for instance garbage collection. The disadvantage of (iii) is, however, there is often a platform-dependent restriction on stack sizes and number of entries, so in practise there are tough frame restrictions, e.g. a maximum offset, which 
should not be exceeded without getting a severe performance penalty on concrete target architectures at the same time. Meyer \cite{lit14} asks to turn garbage collection steadily on during program execution. This implies for efficient execution runtime critical parts will not trigger dynamic memory operations and those operations are opted out as efficiently as it can possibly be done.

Approaches (i) and (ii) are similar, both describe the memory state, although (i) describes the entire dynamic heap as an entire graph, where edges are region dependencies and vertices are locations. The problem with (i) is \textit{locality}, because if a particular function is called, the entire graph has to be specified before, after and during the call, where approach (ii) allows to hide all non-affected heaps (\textit{framed heaps}) – this is what is meant by \textit{locality principle} in terms of \textit{Separation Logic} \cite{lit24}. The most important concept behind Separation Logic \cite{lit24,lit23,lit6} is the specification of two non-interleaving memory regions. Heaps might be composed, and programs may change heaps. If two heaps are connected, then the dependency has to be added explicitly to a current heap's specification. If a heap depends on some other heap data, then this is called \textit{aliasing} (or “\textit{big brother property}” as found in \cite{lit14}).

The first implementation which makes use of Separation Logic is Smallfoot \cite{lit5,lit6}. In order to extend deductive reasoning capabilities, an abductive approach was proposed, called bi-abduction \cite{lit7}, for Separation Logic, which is a constructive guess of unchanged heaps by a greedy symbolic table-construction algorithm that chooses bigger rules first. The extension of Smallfoot is called \textit{SpaceInvader}.

Hurlin extends in \cite{lit10} the classic Separation Logic proposed in \cite{lit23} by classes for a Java-like language. He suggests a \textit{heap factorization}, an attempt to normalise heaps in order to remove redundant heap specification fragments which are considered as \textit{noise}, even if the main goal of his thesis is focused on multi-threaded applications. He re-uses the same concept of \textit{abstract predicates} as it was introduced by \cite{lit18,lit8}, and generates unchanged parts during deduction with a parallel algorithm.

Parkinson \cite{lit18} introduces a Java-like language with object-orientated features. Nevertheless, many problems are not being addressed yet: abstraction mismatch on encapsulation and inheritance, particularly, the problem of expanding specifications in subclasses seems to be a real
hinder in simple and elegant specifications. The most essential contribution of \cite{lit18} is the introduction of \textit{abstract predicates}, although there are currently tough restrictions concerning expressibility. Super calls, static fields, reflection, inner classes and quantified predicates, for example, are currently missing language features.

Verifast \cite{lit11} is another forward verifier based on Separation Logic. In comparison to all previously introduced verifiers which do very similar operations on the heap, all introduced conventions per tool differ strongly, and it does not automatically process loop invariants nor predicates -– here it depends entirely on user-interaction or requires explicit injections within specification annotations inside the program which are used as internal reorganisation commands.\\

In \cite{lit2} objects as class-instances are treated as records, typing and verification rules are introduced and a soundness proof is provided. Problems which neither in \cite{lit2} nor \cite{lit1} are addressed are that objects may have references to other objects and that a lack in abstraction causes a dramatic increase in specification length which makes it in practice impossible to read and understand specification to a full extend. The memory state is specified by temporal predicates and a result-register for the previous computation step's result. There is no general recursive definition allowed, although \cite{lit12} attempts to relax this hard restriction by an algebraic ideal-construction. Still there are hard restrictions, such as no aliasing nor late binding at all, and object-records only which even may become unsound for eager type evaluation.

\section{Architecture and Design}

Before going on with more details on the architecture, the prerequisites on the architecture shall be summarised. The proposed architecture may be considered for teaching purposes in the future:

\begin{enumerate}
 \item Automatic proof. The program and its annotations shall be sufficient in order to get the verification run. If there is an endless cycle in the proof however, there shall be no mandatory recognition, since termination is beyond the main focus of this work.
 \item Openness. The provided architecture shall be open for extensions and variability, and the attached models shall be exportable, so it might eventually be passed through to another arbitrary model transformer, if needed.
 \item Extensibility. The target language shall be fixed but interchangeable with an imperative programming language in the front-end. The rules and user-defined predicates shall be designed amendable, so the user may want to add rules directly to the rule set.
 \item Plausibility: There shall be configurable visualisation facilities, so the incoming annotated program may be retrospected on each stage of the verification process. If, for instance, a proof fails or stops abruptly the user would perhaps like to see the proof tree and a counter-example.
\end{enumerate}

\begin{figure}
\includegraphics[height=6.7cm]{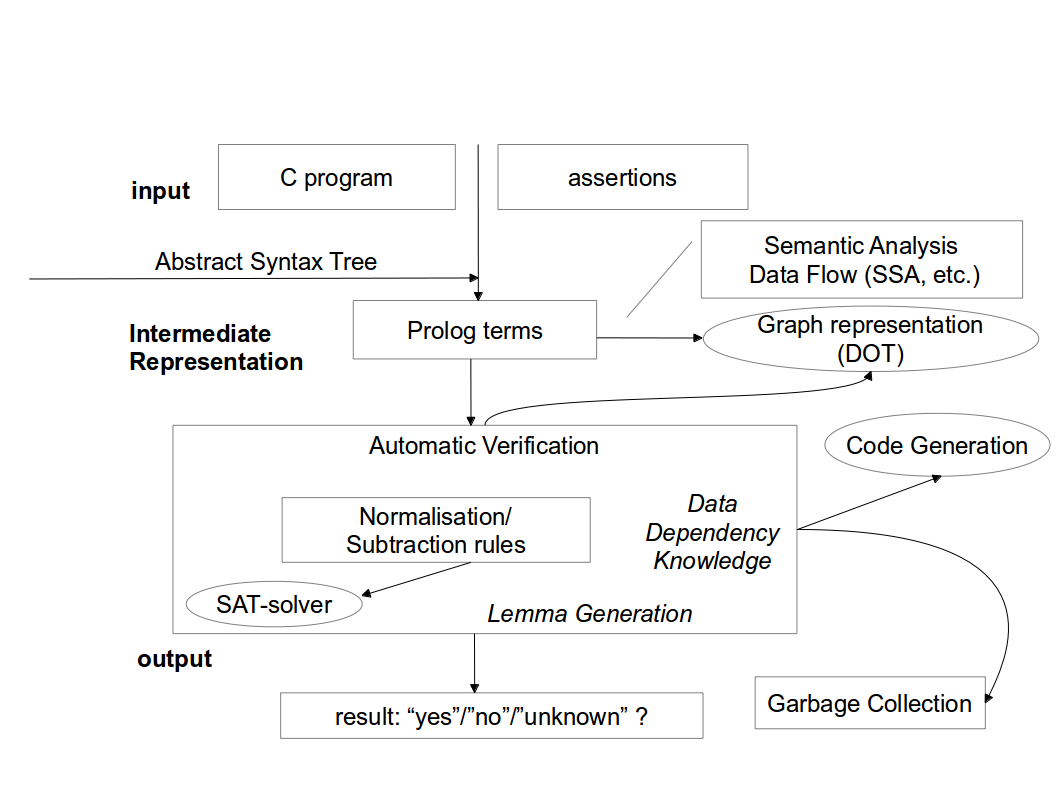}
\caption{Verification architecture for Prolog-based reasoning on dynamic memory}
\label{architectureImg}
\end{figure}

In figure \ref{architectureImg} the architecture of the Prolog-based verification system is shown. The input is a C-program with object-orientated extension that is annotated with assertions specifying the dynamic memory. The shortened syntax can be described by the Extended Backus-Naur form in figure \ref{syntaxCProgram}. Not mentioned definitions as actual parameters, blocks, class methods, variable declarations have been skipped here for the sake of readability and follow mostly ANSI C.  For readability purposes the expression sub-grammar has not been expanded according to its precedence hierarchy nor for optional assertions. \texttt{new} and \texttt{delete} reserve/free new chunks in the heap associated with previously defined locations. The access to heap memory is performed by \texttt{[<location>]}, where \texttt{<location>} denotes either a field variable, another object's field or a either of those with an offset in order specify non-aligned memory regions, for instance. The rule \texttt{<funcall>} 
denotes the syntax for a method call, which may have a object specifier optionally and a method name which is required to exist with the matching total number and types of parameters being passed as expressions.

C-programs are annotated by assertions which are injected as usual Prolog terms into blocks. Blocks are encoded as lists of statement-terms. Assertions are inductively defined and can be found in figure \ref{syntaxAssertions}. Keep in mind the expression might request object references and $\alpha(\vec{p})$” assumes predicate named $\alpha$ was defined prior to using it, and $\vec{p}$ contains as many actual parameters as the arity of predicate $\alpha$ require there are.

\begin{figure}[h]
\begin{grammar}
<prog> ::= <class> <id> '\{' \{ <field> | <method> \} '\}'

<location_1> ::= <id> | <id> '.' <id> | 'this' '.' <id>

<location> ::= <location_1> [ ( '+' | '-' ) <int> ]

<stmt> ::= <lhs> '=' \{ <lhs> '=' \} <expr>
  \alt 'if' <cond> <block> [ 'else' <block> ]
  \alt 'while' <cond> <block>
  \alt 'new' '(' <location_1> ')' 
  \alt 'delete' '(' <location_1> ')'
  \alt <func_call>

<lhs> ::= <location_1> | '[' <location> ']'
  
<cond> ::= <expr> <rel> <expr>

<rel> ::= '\&\&' | '||' | '==' | '!=' | '$\le$' | '$\ge$' | '$>$' | '$\textless$'

<expr> ::= <expr> ( '+' | '-' | '*'  ) <expr> 
\alt '-' <expr>
\alt '[' <location> ']'
\alt [ ( 'this' | <id> ) '.' ] <id> '(' <act_params> ')'
\alt <location>
\alt <int>

<func_call> ::= [ ( 'this' | <id> ) '.' ] <id> '(' <act_params> ')'
\end{grammar}
\caption{Syntax definition of C-programs with object-oriented extension}
\label{syntaxCProgram}
\end{figure}

For example, 

\begin{verbatim}
int f(int a, int b) @ a<10 @ {
  id=2; a=1; b=6;
} @ a->5 * b->c * c->object(myClass1,15) @
\end{verbatim}

is transformed into this Prolog-term:

\begin{verbatim}
function(f, int, 
  [param(a,int), param(b,int)],
  [assert(le(a,10)),
   assign(id,2), assign(a,1), assign(b,6),
   assert(a->5 * b->c * 
          c->object(myClass1,15))]) 
\end{verbatim}
An important specification fragment of the intermediate Prolog-term syntax can be found in figure \ref{syntaxProlog}, where the remaining part is close to the syntax of figure \ref{syntaxCProgram}. Apart from \textit{'ite'} which represents a \textit{if-then-else}-construct with at least one block for the if-case and one more optional block for the else-block -- as long as it was provided, there are also while-loops and further constructs, like class definitions, which will not be mentioned here for simplicity purposes. All expressions, particularly with binary operators, are encoded as terms where the literal operator becomes the functor, for example \texttt{add(i,7)}.

\begin{figure}[h]
$H::=$ 
\begin{tabular}[t]{ll}
$\textbf{emp} \ | \ \textbf{true} \ | \ \textbf{false} \ |$ & \textit{atomic formulae}\\
$x \mapsto E \ |$  & \textit{location map}\\
$H * H \ |$ & \textit{heap separation}\\
$H \vee H \ | \ H \wedge H \ |$ & \textit{conjunction}\\
$\exists x.H \ |$ & \textit{quantification}\\
$\alpha(\vec{p})$ & \textit{predicate unfold}\\
\multicolumn{2}{c}{where}\\
\multicolumn{2}{l}{$x$ is a location}\\
\multicolumn{2}{l}{$E$ is a well-defined expression (enumeration)}\\
\multicolumn{2}{l}{$\vec{p}$ is a comma-separated parameter vector}
\end{tabular}
\caption{Syntax definition of heap and stack assertions}
\label{syntaxAssertions}
\end{figure}

\begin{figure}[b]
\begin{grammar}
<stm> ::= ... | 'new' '(' <loc_1> ')' 
 \alt 'delete' '(' <loc_1> ')'
 \alt 'funcall' '(' <id> [ ',' <act_params> ] ) | ...
 \alt 'ite' '(' <cond> ',' <block> [ ',' <block> ] ')'

<loc_1> ::= <id> | 'oa' '(' <id> '.' <id> ')'

<loc> ::= 'offset' '(' <loc_1> [ ',' <offset> ] ')'

<offset> ::= <int> | 'minus' '(' '0' ',' <int> ')'

<expr> ::= ( 'add' | 'sub' | 'mul' ) '(' <expr> ',' <expr> ')'
 \alt 'mem' '(' <loc> ')'
 \alt <loc> | <int>
 \alt 'funcall' '(' <id> [ ',' <act_params> ] ')'
\end{grammar}
\caption{Syntax definition of Prolog-terms}
\label{syntaxProlog}
\end{figure}

Since the architecture is designed flexible, it allows the user to interchange the compiler front-end for a different language, so the user has the possibility to write own Prolog-terms directly without even having an ordinary C-program. In this case syntax and semantic constraints remain on full responsibility to the user. Prolog-terms are internally checked and may also be directed to a graphical output, e.g. for proof tree visualisation. Antlr 4 \cite{lit19} is currently used as compilation front-end.\\

Once the Prolog term is constructed, it can be passed to the verification. Hereby, the term is now processed while the internal environment, which has to keep the states of the memory, needs to be updated after every statement. All locals are residing in stack, where dynamically allocated memory locations may remain in memory –-- even if a stack-based variable stores a dynamic address it would be freed at the end of a block.

If we decide to specify a list concatenation of two lists, we have several opportunities to describe the heap. If \texttt{list(s,e)} denotes a heap predicate where \texttt{s} is the location of the beginning root element of a list, and \texttt{e} denotes the last element in that list, then having two lists \texttt{x} and \texttt{y} with \texttt{x->a,b,c} and \texttt{y->d,e,f} will concatenate for instance to either (i) \texttt{x->a,b,c,d,e,f * y->f} or to (ii) \texttt{x->a,b,c * y->d,e,f * z->a,b,c,d,e,f}. Remark: The ','-operator is defined as a list constructor with variable input amount for all consecutive objects currently in memory linked together to a simply-linked list \cite{lit23}. The main difference between (i) and (ii) is that (i) requires only a single assignment if the end of \texttt{x} is known, therefore \texttt{x} and \texttt{y} are no more as they used to be before concatenation. (ii) creates an entirely new copy of all element from \texttt{x} and \texttt{y} and does not touch neither \texttt{x} nor \texttt{y}. (ii) is safer from a 
general reuse perspective, but it is considerably slower and consumes more memory due to additional copies to be generated.

Finally, the SMT-solver is required whenever taking out trivial calculations, for instance in basic arithmetics. For instance, if there is an expression that might be reduced to a value, then this should in general be tried first before triggering a certain rule. Beside finding solutions to basic arithmetic and other theories, re-arrangement needs to be taken into consideration. Formal rules will usually also not deal too much about heap permutation, although a strategy must be found and is crucial in fact for the overall performance. 

\section{Conclusion}
So far the open architecture was presented providing several suggestions for further research activities. Prolog was proposed as specification and proof platform for memory-specific research, e.g. on extending the expressibility of abstract predicates or abduction. We believe, questions related to abduction in Separation Logic with objects still have not been profoundly investigated yet, as well as some object-oriented features like polymorphism in Separation Logic.

The platform might be used to incorporate with existing compiler packages in order to research improvement on code optimization during the alias analysis phase, but also garbage collection, based on knowledge obtained during the dynamic memory verification. 

Further rules of normalisation and re-arrangement will be applied to cover more real world scenarios, particularly in order to resolve arithmetic equivalency by the integration of a SMT-solver (\cite{lit17,lit22}).

Related work includes Jacobs \cite{lit11} who suggests to investigate Banerjee's Regional Logic approach \cite{lit4} as substitute for the Symbolic Execution approach \cite{lit6}.

\newpage



\begin{thebibliography}{1}
\bibitem{lit1}
 Abadi M., Cardelli L. \emph{A. Theory of Objects}. New York: Springer, 1996, 396 p.
\bibitem{lit2}
 Abadi M., Leino K. R. M. \emph{A Logic of Object-Oriented Programs}, Proc. of the 7th Int. Joint Conf. CAAP/FASE on Theory and Practice of Software Development, Springer, 1997, pp. 682-696.
\bibitem{lit3}
 Allen R., Kennedy K. \emph{Optimizing Compilers for Modern Architectures}. 2001, 790p.
\bibitem{lit4}
 Banerjee A., Naumann D. A. and Rosenberg S. \emph{Regional logic for local reasoning about global invariants}. ECOOP, LNCS 5142, 2008, pp. 387-411 
\bibitem{lit5}
 Berdine J., Calcagno C. and O'Hearn P. W. \emph{Smallfoot: Modular Automatic Assertion Checking with Separation Logic}. FMCO, 2005, pp. 115-137.
\bibitem{lit6}
 Berdine J., Calcagno C. and O'Hearn P. W. \emph{Symbolic Execution with Separation Logic}. APLAS, 2005, pp. 52-68.
\bibitem{lit7}
 Calcagno C., Distefano D., O'Hearn P. and Yang H. \emph{Compositional shape analysis by means of bi-abduction}. Proceedings of the 36th annual ACM SIGPLAN-SIGACT symposium on Principles of programming languages, 2009, 36, pp. 289-300 .
\bibitem{lit8}
 Distefano D., Parkinson M. \emph{jStar: Towards practical verification for Java}. OOPSLA, 2008, pp. 213-226.
\bibitem{lit9}
 Hind M., \emph{Pointer Analysis: Haven't We Solved This Problem Yet?} ACM, PASTE'01, 2001, pp. 54-61.
\bibitem{lit10}
 Hurlin C. \emph{Specification and Verification of Multithreaded Object-Oriented Programs with Separation Logic}. PhD Thesis, Université Nice - Sophia Antipolis, 2009, 195p.
\bibitem{lit11}
 Jacobs B., Piessens F. \emph{The VeriFast Program Verifier}. Leuven University, 2008, 5p.
\bibitem{lit21}
 Koch H., Schenkel A., Wittwer P. \emph{Computer-assisted Proofs in Analysis and Programming in Logic: A Case Study}. SIAM Review, 1996, no.4(38), pp. 565-604.
\bibitem{lit12}
 Leino K. R. M. \emph{Recursive object types in a logic of object-oriented programs}. Nordic J. of Computing, (5), 1998, pp. 330-360.
\bibitem{lit13}
 Meyer B. \emph{Proving Pointer Program Properties - Part 1: Context and Overview}, Journal of Object Technology, no.2(2), March-April 2003, pp. 87-108.
\bibitem{lit14}
 Meyer B. \emph{Proving Pointer Program Properties - Part 2: The Overall Object Structure}, Journal of Object Technology, no.3(2), May-June 2003, pp. 77-100.
\bibitem{lit15}
 Miller B. P., Fredriksen L. and So B. \emph{An Empirical Study of the Reliability of UNIX Utilities}. Proc. of the Workshop of Parallel and Distributed Debugging, Digital Equipment Corporation, 1990, pp. 1-22.
\bibitem{lit16}
 Muchnik S. \emph{Advanced Compiler Design and Implementation}. Morgan Kaufman, 2007, 856p.
\bibitem{lit17}
 Nanevski A., Morrisett G., Shinnar A., Govereau P. and Birkedal L. \emph{Ynot: Reasoning with the awkward squad}, ACM SIGPLAN Int. Conf. on Functional Programming, 2008, p. 12.
\bibitem{lit22}
 OpenSMT project. http://code.google.com/p/opensmt/
\bibitem{lit18}
 Parkinson M. \emph{Local Reasoning for Java}. PhD Thesis, Cambridge University, 2005, 169 p.
\bibitem{lit19}
 Parr T. \emph{The Definitive ANTLR 4 Reference: Building Domain-Specific Languages}. O'Reilly, 2013, 328p.
\bibitem{lit20}
 Pavlu V. \emph{Shape-based Alias Analysis. Computing Alias Sets from Shape Graphs to Evaluate the Precision of Shape Analyses}. VDM Verlag Dr. M\"{u}ller, 2010, 117p.
\bibitem{lit23}
 Reynolds J. C. \emph{Separation Logic: A Logic for Shared Mutable Data Structures}, Lecture Notes in Computer Science, 2002, pp.55-74.
\bibitem{lit24}
 Reynolds J. C. \emph{An Introduction to Separation Logic}. Carnegie Mellon University, 2009, 204p.
\bibitem{lit25}
 SAFECode within LLVM project. http://llvm.org
\bibitem{lit26}
 Sagiv M., Reps T., Wilhelm R. \emph{Parametric shape analysis via 3-valued logic}. ACM Trans. Program. Lang. Syst., 2002, 24, pp. 217-298.
\bibitem{lit27}
 Sterling L., Shapiro E. \emph{The Art of Prolog (2nd ed.): Advanced Programming Techniques}, MIT Press, 1994, 552 p.
\bibitem{lit28}
 Tofte M., Talpin J.-P. \emph{Implementation of the typed call-by-value $\lambda$-calculus using a stack of regions}. Proc. of the 21st ACM SIGPLAN-SIGACT. 1994, pp. 188-201.
\bibitem{lit29}
 Valgrind project. http://www.valgrind.org 
\end{thebibliography}

\end{document}